\documentstyle[12pt,epsf]{article}

\newcommand{\lsim}{\mathrel{\lower4pt\hbox{$\sim$}}
\hskip-12.5pt\raise1.6pt\hbox{$<$}\;}

\newcommand{\gsim}{\mathrel{\lower4pt\hbox{$\sim$}}
\hskip-12.5pt\raise1.6pt\hbox{$>$}\;}

\begin{document}

\begin{flushright}
JLAB-TH-96-14 \\
BNL-~~~~~~~~~~~~~~~
\end{flushright}
\bigskip
\begin{center}
{\bf CP Conserving and CP Violating Asymmetries in $e^+e^-\to t\bar
t\nu_e\bar\nu_e$} \\
{\bf at the NLC: Application to Determining the Higgs Width}\\
\bigskip\bigskip
David Atwood$^a$ and Amarjit Soni$^b$
\end{center}

\bigskip

\begin{flushleft}
$a$) Theory Group, CEBAF, Newport News, VA\ \ 23606 \\
$b$) Theory Group, Brookhaven National Laboratory, Upton, NY\ \ 11973
\end{flushleft}
\bigskip

\begin{quote}
{\bf Abstract}: The polarization asymmetries of $t$, $\bar t$ quarks
can be used in the reaction $e^+e^-\to t\bar t\nu_e\bar\nu_e$ to
measure the Higgs width, in the Standard Model or in its extensions,
and to search for non-standard CP violating phases. As an application of
the CPT theorem the Higgs width is
monitored through a CP-even, $T_N$-odd, polarization asymmetry, $B_y$.
CP violation manifests through interference at tree graph level with the
resonant Higgs amplitude. Consequently, the asymmetries are all quite sizable
and can be in the range of 10--50\% for a wide choice of the Higgs mass.
\end{quote}
\bigskip

A high energy $e^+e^-$ collider, called the Next Linear Collider (NLC),
with center of mass (cm) energies ranging from $\sqrt{s}=0.5$--1.5 TeV has
been receiving considerable attention in the last few years
\cite{epemcol}--\cite{muryama}. The clean environment that it should possess
endows it
with a unique ability to probe detailed issues pertaining to dynamics
of symmetry breaking, flavor violations, CP nonconservation etc. Thus
such a facility should nicely complement the physics reach of the Large
Hadron Collider (LHC).

In this work we discuss how the reaction

\begin{equation}
e^++e^-\to t\bar t\nu_e\bar\nu_e \label{epem}
\end{equation}

\noindent can be used for confronting two important issues, namely,
extracting the width of the Higgs particle(s) and searching for
non-standard CP violating phases. The key point is that due to its
large mass the top quark does not bind into hadrons \cite{bigi}. Decays
of the top quark then act as very effective analyzer of its spin
\cite{schmidt}. This ability to track the top spin allows us to
construct CP-even and CP-odd observables utilizing the top
polarization. CP even observables that are odd under naive time
reversal \cite{selfcont} ($T_N$) have the important property, that
follows from the CPT theorem of Quantum Field Theory, that they are
driven by the absorptive part of Feynman amplitudes. In the process
under consideration, i.e.\ reaction (\ref{epem}), the absorptive part of
the amplitude is proportional to the width of the Higgs particle thus
allowing the possibility of experimentally measuring the Higgs width.
Since the Higgs width is an important characteristic of the nature of
the electroweak symmetry breaking mechanism its measurement is clearly
significant.
In addition, of course, CP odd observables, utilizing the top spin can
be readily constructed to enable us to probe the presence of
non-standard CP-violating phases that reside in the neutral Higgs
sector \cite{bernreuther} in extensions of the Standard Model (SM), say in
a  two Higgs doublet model (2HDM).

A distinctive feature of reaction (\ref{epem}) that we exploit is that CP
violation can manifest itself by interference of the Higgs resonance in tree
amplitudes. The resulting asymmetries are substantial \cite{cpasym,cpvioeff}.
The CP violating
asymmetries that are of interest can be in the range of tens of
percents, quite often 10--50\%. The CP conserving, $T_N$-odd asymmetry can
also be
as big as 40\%. This is especially striking as this asymmetry  is
a feature of the SM itself and does not require non-standard physics; it
has a very important application to determining the Higgs width.
Of course,  measurements of the Higgs width will serve to
elucidate whether the Higgs is standard or not.

For convenience we work in the analog of the equivalent photon
approximation, i.e.\ the equivalent $W$-boson approximation. At large
c.m.\ energies i.e.\ as $s/M^2_W$ becomes very large, $s$ being the total
energy squared in the $e^+e^-$ c.m.\ frame, the cross section
for reaction (\ref{epem}) is dominated by the collisions of the
longitudinally polarized $W$'s \cite{cahn}. In this approximation
reaction (\ref{epem}) can be replaced by a simpler reaction, i.e.\ the
$W$-boson fusion process:

\begin{equation}
W^++W^- \to t +\bar t \label{wpwm}
\end{equation}

\noindent Indeed, the salient features of the underlying physics can be
succinctly stated in the context of reaction (\ref{wpwm}). Furthermore,
although the expression for the cross section thus deduced is accurate
only in the leading log (in $s/M^2_W$) approximation, the asymmetries
that are of more central importance to this work hold to a much better
accuracy as they result from the ratios of cross-sections.

To lowest order  there are four Feynman graphs, shown in Fig.~1,
relevant to reaction (\ref{wpwm}). The blob in Fig.~1(a) indicates that
the propagator of the Higgs resonance is
highly unstable i.e.\ it possesses a non-negligible width which can
in fact be a substantial fraction of its mass depending on what the mass
is. It is this Breit-Wigner nature of the scalar propagator in Fig.~1
that endows the Feynman amplitudes for reactions (\ref{epem}) or
(\ref{wpwm}) to have an absorptive part that is of special interest to
us \cite{cpcon}.

Consider now the $t$, $\bar t$ polarization asymmetries.
In the rest frame of the t let us define the basis vectors:
$-e_z\propto (\vec p_{W+}+\vec p_{W-})$;
$e_y\propto \vec p_{W+}\times \vec p_{W-}$ and
$e_x=e_y\times e_z$. Let $P_j$ (for $j=x,y$ or $z$) be the
polarization of the $t$ along $e_x$, $e_y$, $e_z$.
For the anti-top we use a similar set of definitions in the
$\bar t$ frame related by charge conjugation:
$-\bar e_z\propto (\vec p_{W-}+\vec p_{W+})$;
$\bar e_y\propto \vec p_{W-}\times \vec p_{W+}$ and
and
$\bar e_x=\bar e_y\times \bar e_z$.
Similarly $\bar P_j$
the polarization of the $\bar t$ is in the $\bar e_x$, $\bar e_y$,
$\bar e_z$
direction.
Combining
the information from $t$, $\bar t$ system we may define the following
asymmetries:

\begin{eqnarray}
A_x = \frac{1}{2} (P_x+\bar P_x) & ; & B_x = \frac{1}{2} (P_x-\bar P_x)
\nonumber \\
A_y = \frac{1}{2} (P_y-\bar P_y) & ; & B_y = \frac{1}{2} (P_y+\bar P_y)
\label{axyz} \\
A_z = \frac{1}{2} (P_z+\bar P_z) & ; & B_z = \frac{1}{2} (P_z-\bar P_z)
\nonumber
\end{eqnarray}

\noindent Here $A$'s are CP-odd and $B$'s are CP-even, also $\{A_x,
B_y,A_z\}$ are CP$T_N$-odd whereas $\{B_x, A_y, B_z\}$ are CP$T_N$-even.
Thus $\{A_x, B_y, A_z\}$ are proportional to the absorptive part of the
Feynman amplitude which receives dominant contribution from the
Higgs  exchange graph of Fig.~1a, particularly if $m_H$ is large. The
other two CP even observables, $B_x$ and $B_z$, are not relevant to our
discussion as they do not receive contributions from an absorptive
phase. Of course, all the CP odd observables are important; they all
require CP violating phase(s) in the underlying
Lagrangian.
$A_y$
is driven by the real part of Feynman amplitudes and
$A_x$
and $A_z$
again, by the imaginary part and thus proportional to the Higgs width.

It is important to stress that since $B_y$ is CP conserving it can be
used for determining the width of Higgs particle(s) in the SM as well as
its extensions. Indeed it should be useful for scalar as well as pseudoscalar
Higgs particles.

Let us now consider, in some generality, models of CP violation based on
the use of an extended Higgs sector.
As is well known such models require
at least two Higgs doublets \cite{bernreuther}.
A feature of
many such models
which has important bearing on the phenomenological
implications of CP violation
is that when the masses of all the
Higgs states are degenerate then CP violation
effects due to the Higgs sector must vanish. This means that in high energy
processes, such as $W^++W^-\to t\bar t$, as the c.m. energy becomes much
larger than the masses of all the Higgs particles, then the total contribution
to CP violation from all the Higgs must necessarily vanish.
For instance the 2HDM in \cite{bernreuther}
has this feature.

In order to produce sample numerical calculations for CP violating effects
where this type of cancelation is built in on the one hand while
on the other hand the number of
free parameters to be considered is
still small,
we will assume
that there are n neutral scalars
and that k of them are degenerate with mass $m_H$ while
the remaining $n-k$ are degenerate with mass
$m_H^\prime$ where $m_H<m_H^\prime$.
For perturbation theory
to remain valid we must also require that $m_{H^\prime}\lsim1$ TeV\null.
Within the states which are degenerate at $m_H$ one can in general perform an
orthogonal rotation so that only one has a coupling
to the top quark
term proportional to $\bar t i \gamma_5 t$. We will denote this Higgs state
$H$.
Likewise one may perform an orthogonal rotation among the $n-k$ states with
mass $m_H^\prime$ so that only one of these states has a coupling proportional
to
$\bar t i \gamma_5 t$ which we will denote $H^\prime$. The remaining $n-2$
Higgs states in the model,
denoted by
$h_i$ for $i=1  \to (n-2)$,
have CP conserving interactions.
All of the CP violation relevant to $WW\to t\bar t$
is thus controlled by $H$ and $H^\prime$.

The Lagrangian terms involving the Higgs coupling to $t\bar t$ and $WW$
can be written as:
\begin{eqnarray}
{\cal L}_{Htt} \!=\!
-{g_W\over 2} {m_t\over m_W} H
\bar t (a_H+ib_H\gamma^5) t
& & \!\!\!\!\!
{\cal L}_{HWW}
\! = \!
g_W c_H m_W g^{\mu\nu} H  W^+_\mu W^-_\nu
\nonumber\\
{\cal L}_{Htt}^\prime \! = \!
-{g_W\over 2} {m_t\over m_W}  H^\prime
\bar t (a_H^\prime+ib_H^\prime\gamma^5) t
& & \!\!\!\!\!
{\cal L}_{HWW}^\prime
\! = \!
g_W c_H^\prime m_W g^{\mu\nu} H^\prime  W^+_\mu W^-_\nu
\nonumber\\
{\cal L}_{h(i)tt} \! = \!
-{g_W\over 2} {m_t\over m_W}  h_i
\bar t a_{h(i)} t
& & \!\!\!\!\!
{\cal L}_{h(i)WW}
\! = \!
g_W c_{h(i)}^\prime m_W g^{\mu\nu} h_i  W^+_\mu W^-_\nu
\label{lag}
\end{eqnarray}

\noindent
For the cancelation to apply
the above couplings are subject to the constraints that
$b_Hc_H+b_H^\prime c_H^\prime=0$.
The corresponding Feynman rules can be easily derived and the
amplitudes and expectation values for the observables of interest
$\{A_x, A_y, A_z, B_y\}$ can be calculated in the standard manner. From
Eqn.~(\ref{lag}) we see, as is also well known, that the $Htt$ vertex
violates CP due to the simultaneous presence of the scalar and
pseudoscalar interactions.

Specifically in
the 2HDM in
\cite{bernreuther}
the  neutral Higgs states $\phi_j$ for $j=1-3$ have couplings
\begin{equation}
a_j=d_{2j}/\sin\beta\ \ \ \
b_j=-d_{3j}/\tan\beta\ \ \ \
c_j=d_{1j}\cos\beta+d_{2j}\sin\beta
\end{equation}
where $d_{ij}$ forms an orthogonal $3\times 3$ matrix.
For our numerical examples
with CP violation
we will identify $H=\phi_1$, $H^\prime=\phi_2$, $h_1=\phi_3$
(where we let $h_1$ and $H^\prime$ be degenerate so $k=1$)
and $d_{31}=-d_{12}=\sin\beta^\prime$;
$d_{32}=d_{11}=-\cos\beta^\prime$;
$d_{23}=1$ with the other components being $0$.
In addition we will use the value $\tan\beta=1/2$
and set $\beta^\prime=\beta$ \cite{beta1}.

We compute the asymmetries as a function of $\sqrt{\hat s}$ for
various $\Gamma_H, m_H$. Note that, we assume that
$\Gamma_H$, $\Gamma_H^\prime$
should be mostly given by decays to $t\bar t$, $WW$ and $ZZ$:

\begin{equation}
\Gamma_H\simeq \Gamma\equiv \Gamma_{H\to t\bar t} +
\Gamma_{H\to WW} + \Gamma_{H\to
ZZ} \label{ttwwzz}
\end{equation}

\noindent To allow for the presence of modes other than the above three
we express the width as:

\begin{equation}
\Gamma_H = \lambda_H \Gamma
\label{phah}
\end{equation}

We now present some of the numerical results. Figs.~2 and 3 show the four
asymmetries of interest to us. Fig.~2 shows the asymmetries as a function
of $\sqrt{\hat{s}}$ for a fixed $m_H$ whereas in Fig.~3 they are shown
as a function of $m_H$ for a fixed $s$. As an illustration
the Higgs masses are held fixed in Fig.~2 at
$m_H=500 GeV$ and $m_H^\prime=1000GeV$
and we use the couplings
with $\tan\beta$ in the version of the model of \cite{bernreuther}
described above.
We
also show the CP conserving $T_N$-odd asymmetry that occurs in the SM,
i.e.\ $B^{\rm SM}_y$. For this purpose we, of course, use the SM
couplings, $a_H=c_H=1$, $b_H=0$.

From Fig.~2 we see that the asymmetries tend to be fairly large ranging from
about 10  to 50\% for a wide range of values of $\sqrt{\hat{s}}$.  The CP
conserving, $T_N$-odd asymmetry, $B_y$, that is
proportional to the Higgs width is also appreciable.
For the SM it can be as big
as about 40\%.

In the leading log approximation that we are using the cross section
for reaction (\ref{epem}) and the corresponding asymmetries can be
readily calculated in the $e^+e^-$ cm frame. Fig.~3 shows the four
relevant asymmetries for $\sqrt{s}=1.5$ TeV;  we fix $m_H^\prime=1000GeV$.
We see that for a wide
range of Higgs masses the asymmetries are appreciable. Indeed $A_z$
approaches 25\% for $m_H\sim500$--700 GeV whereas $A_y$ tends to be large,
i.e.\ around 30\%, when $m_H$ is in the range 100--300 GeV\null.
For the SM, the CP
conserving asymmetry ($B^{\rm SM}_y$) is around 10--30\% for $m_H\gsim100$
GeV\null.

Even with an ideal detector it is, of course, not possible to measure the
polarization
of a top quark with 100\% precision. Let us here consider two possible modes
(and their conjugates in the case of $\bar t$)
useful for
polarimetry:

\begin{enumerate}
\item
The decay $t\to W^+b$ with $W^+\to \ell^+\nu$, where $\ell= e$, $\mu$. In this
case we will include only the hadronic decays of the  $\bar t$  to avoid
problems in reconstruction.
\item
The decay $t\to W^+b$ with $W^+\to {\rm hadrons}$. Now we exclude the
decay of $\bar t$ to a $\tau^-$ to avoid problems due to
reconstruction.
\end{enumerate}

Case (1) occurs with a branching ratio of about $B_1=(2/9)(2/3)=4/27$
where $(2/9)$ is the probability that the $t$ decays to $b e\nu$ or $b\mu\nu$
and
$(2/3)$ is the probability that the $\bar t $ decays hadronically.
If the top quark has polarization $P$ in a given direction then, in general,
the angular distribution of the lepton is
$\propto (1+R_1 P \cos\eta_l)$
where $\eta_l$ is the
angle between the polarization axis and the lepton momentum in the top
frame. For top decays in the SM, $R_1=1$. The optimal
(in the sense that it minimizes statistical error)
method to obtain the value of $P$ is to use
$P^{(1)}=3<\cos\eta_l>/R_1$ where $P^{(1)}$ therefore is the polarization
extracted using this method.

Likewise in case (2) the distribution of the $W$ momentum in the top frame is
$\propto (1+R_2 P \cos\eta_W)$
where, in the SM,  $R_2= (m_t^2-2m_W^2)/(m_t^2+2m_W^2)$ and in this case
$B_2=(2/3)(8/9)=16/27$;
where (2/3) is the probability that the $t$ decays hadronically and
(8/9) is the probability that the $\bar t$ does not decay to a $\tau$.
Here $\eta_W$ is the angle
between the momentum of the $W$ and the polarization
axis.
To obtain the polarization from this mode we can use
$P^{(2)}=3<\cos\eta_W>/R_2$.

The above two cases may be combined to obtain
$P^{12}= (B_1R_1^2P^{(1)}+B_2R_2^2P^{(2)})/$ $(B_1R_1^2+B_2R_2^2)$.
Bearing in mind that the asymmetries $A_i, B_i$
are combinations of the $t$ and $\bar t$ polarizations and that
in each event
we can potentially measure
the polarization of a $t$ and a $\bar t$, the number of events
needed to obtain a 3-$\sigma$ signal is
\begin{equation}
N_{t\bar t}^{3\sigma}
=
{27\over 2}
(R_1^2 B_1 + R_2^2B_2)^{-1} a^{-2}
\end{equation}
where $a$ is the asymmetry in question (either $A_i$ or $B_i$). Numerically
then
$N^{3\sigma}\approx 52 a^{-2}$ requiring some 5200 events for an asymmetry
of 10\%.

Fig.~4 shows the cross-section as a function of $m_H$ for $\sqrt{s}
=1.5$ TeV\null. We see that the typical cross-section tends to be a few
$fb$. For example, for $m_H=500$ GeV, it is around $5fb$ for the SM and
can be about 15 $fb$ in a 2HDM with
the couplings described above.
At
$\sqrt{s}=1.5$ TeV the projected luminosity is about $5\times10^{34}$
cm$^{-2}$ s$^{-1}$ \cite{epemcol}--\cite{miyamoto}. Thus a cross section
of $10fb$ would yield about
5000 events rendering it feasible to detect asymmetries $\gsim 10\%$.

To summarize, the reaction $e^+e^-\to \nu_e\bar \nu_e t\bar t$ can be a
very powerful probe of CP violation.
In models with an extended Higgs sector
appreciable asymmetries result through interference of the Higgs
resonance with tree graph amplitudes. It can also be very useful for
extracting the Higgs width in the Standard Model or
in its extensions.

\bigskip\bigskip

This research was supported in part by the U.S. DOE contracts
DC-AC05-84ER40150 (CEBAF) and DE-AC-76CH00016 (BNL).

\newpage

\begin{center}
{\Large\bf
Figure Captions
}
\end{center}

Figure 1:
The Feynman diagrams that participate in the sub-process
$W^+W^-\to t\bar t$. The blob in Fig.~1a represents the width of the
Higgs resonance and the cut across the blob is to indicate the
imaginary part.

\bigskip

Figure 2:
A graph of the asymmetries $A_x$ (solid); $A_y$ (dashes);
and $A_z$ (dots)
as a function of $\sqrt{\hat s}$ given
$m_H=500$~GeV
$m_H^\prime=1000$~GeV
and the coupling parameters
for $\tan\beta=1/2$ as described in the text.
The  dash-dot curve is the asymmetry
$B^{\rm SM}_y$ for the standard model couplings
$a_H=c_H=\lambda_H=1$ and $b_H=0$ with no $H^\prime$ present.

\bigskip
Figure 3:
A graph of the asymmetries integrated over $\hat s$
as a function of
$m_H$ for $\sqrt{s}=1500$~GeV and $m_H^\prime=1000$~GeV.
See Fig.~2 for notations.

\bigskip
Figure 4:
Cross section (in picobarns) as a
function of $m_H$ for $\sqrt{s}=1.5$ TeV\null. Solid
is for SM with $a_H=1=c_H$, $b_H=0$, dashed is for 2HDM with
the couplings described in the text.
The dotted line is for a 2HDM with
$a_H=-1$, $c_H=1$; $b_H=0$ while $C_H^\prime=0$.
For all three cases $\lambda_H,\lambda^\prime_H=1$ is assumed.

\newpage
\begin{figure}
\epsfxsize=6.in
\epsffile{wt_f1.eps}
\end{figure}

\newpage
\begin{figure}
\epsfxsize=6.in
\epsffile{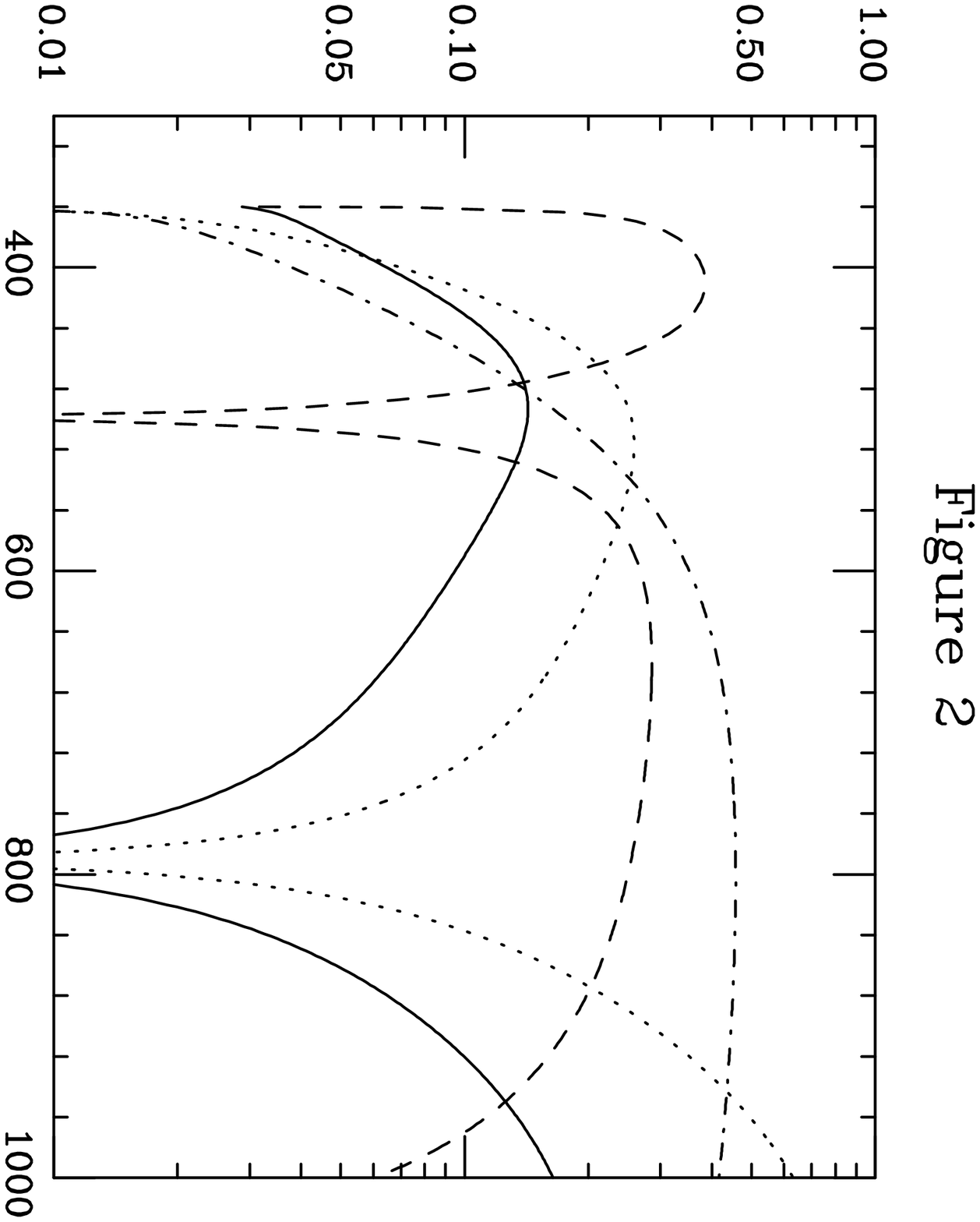}
\end{figure}

\newpage
\begin{figure}
\epsfxsize=6.in
\epsffile{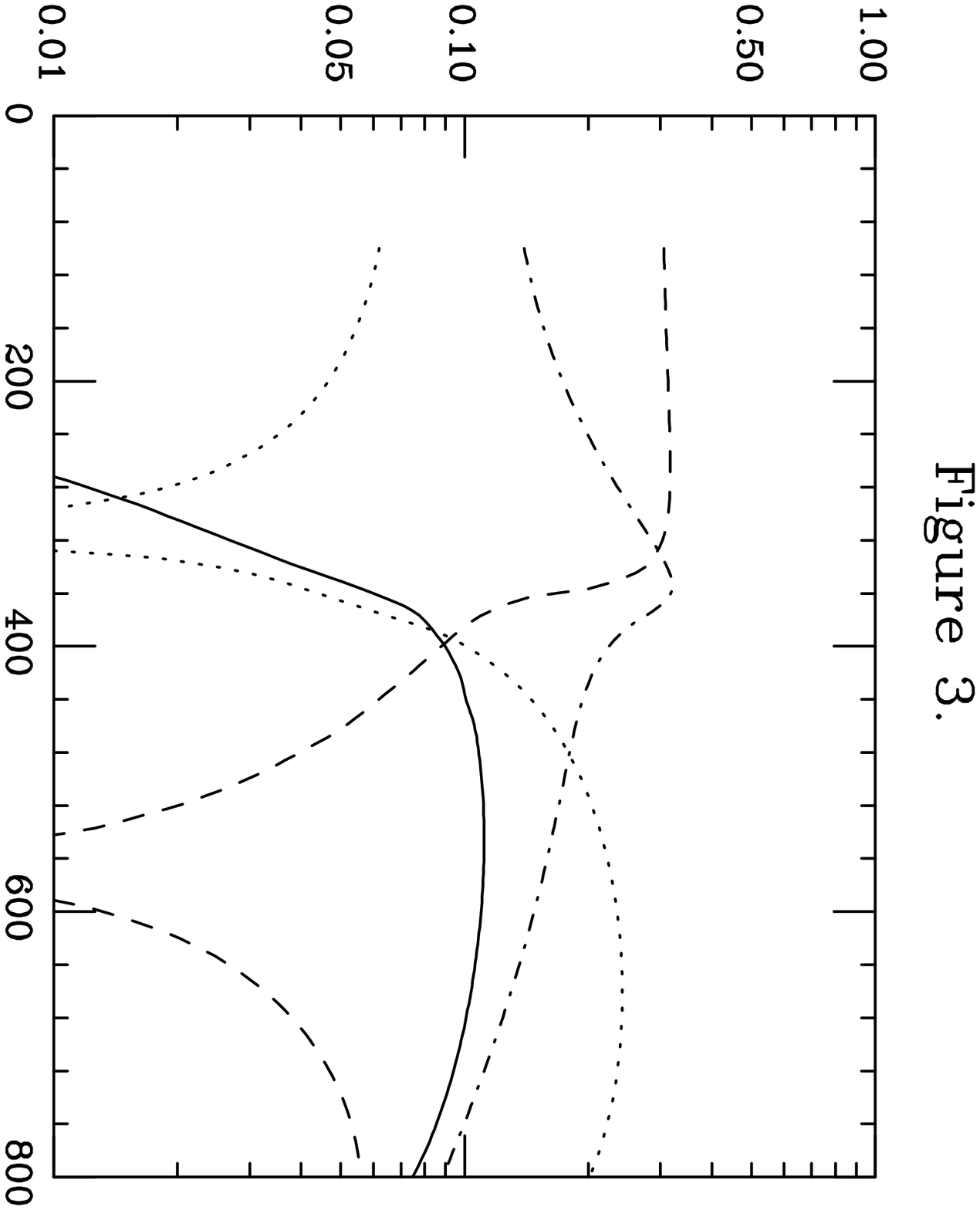}
\end{figure}

\newpage
\begin{figure}
\epsfxsize=6.in
\epsffile{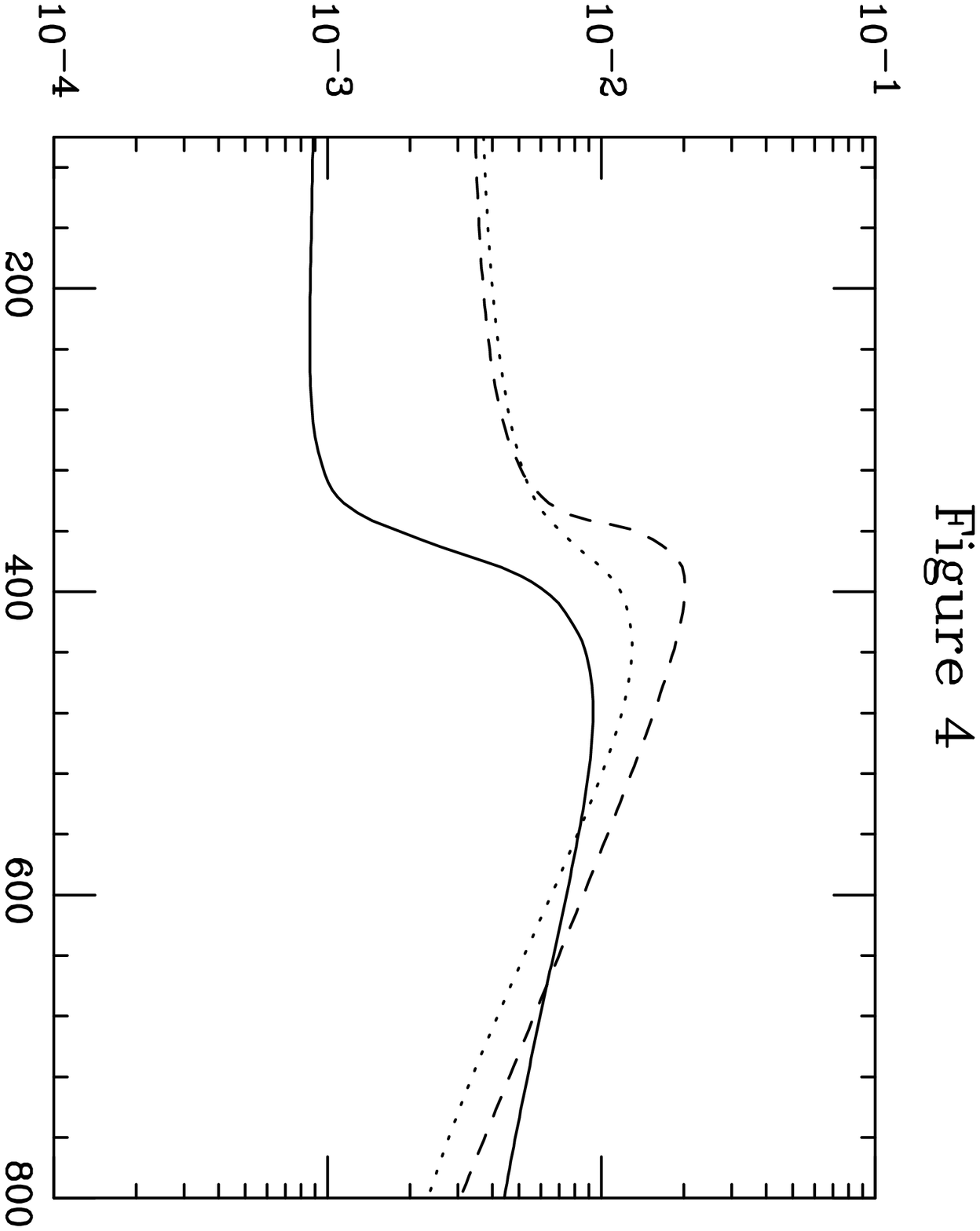}
\end{figure}


\begin{thebibliography}{99}
\bibitem{selfcont} For the purpose of self-containment we should
mention that $T_N$ (naive time reversal) differs from the true time
reversal operator in that the latter requires switching of initial and
final states which the former does not.
\bibitem{cpvioeff} Throughout this paper CP violation effects due to
the CKM phase of the SM will be neglected as they are expected to be
extremely small.
\bibitem{cpcon} CP conserving absorptive parts can, of course, also
arise through radiative corrections of Figs.~1b and 1c even in the absence
of the Higgs
exchange graph in Fig.~1a. For example, gluon exchanges will render the
$WW\to t\bar t$ amplitude complex at one loop. We are taking the view
that these provide a calculable, and most likely, subdominant
correction to the dominant effect, namely due to the Higgs width,
especially for $m_H\gsim100$ GeV, that is our primary concern here.
\bibitem{beta1} There is little change in the asymmetries
if tan $\beta=1$ is used instead.
\end{thebibliography}
\end{document}